\documentclass[a4paper]{INTERSPEECH2024}

\usepackage{graphicx}
\usepackage{amssymb,amsmath,bm}
\usepackage{textcomp}
\usepackage{url}
\usepackage[utf8]{inputenc}
\usepackage{CJKutf8}
\usepackage{multirow}
\usepackage{comment}
\usepackage{subfigure}

\newcolumntype{P}[1]{>{\centering\arraybackslash}p{#1}}

\interspeechcameraready

\ninept

\title{SRC4VC: Smartphone-Recorded Corpus for Voice Conversion Benchmark}

\name[affiliation={1}]{Yuki}{Saito}
\name[affiliation={1}]{Takuto}{Igarashi}
\name[affiliation={1}]{Kentaro}{Seki}
\name[affiliation={1,2}]{Shinnosuke}{Takamichi}
\name[affiliation={3}]{Ryuichi}{Yamamoto}
\name[affiliation={3}]{\\Kentaro}{Tachibana}
\name[affiliation={1}]{Hiroshi}{Saruwatari}

\address{
$^1$The University of Tokyo, Japan,
$^2$Keio University Japan.
$^3$LY Corp., Japan.
}
\email{
yuuki\_saito@ipc.i.u-tokyo.ac.jp
}

\begin{document}
\setlength{\abovedisplayskip}{5pt} 
\setlength{\belowdisplayskip}{5pt} 
\setlength\floatsep{6pt} 
\setlength\intextsep{6pt} 
\setlength\textfloatsep{6pt} 
\setlength{\dbltextfloatsep}{6pt} 
\setlength{\dblfloatsep}{5pt}

\maketitle

\begin{abstract}

We present {\it SRC4VC}, a new corpus containing 11 hours of speech recorded on smartphones by 100 Japanese speakers. Although high-quality multi-speaker corpora can advance voice conversion (VC) technologies, they are not always suitable for testing VC when low-quality speech recording is given as the input. To this end, we first asked 100 crowdworkers to record their voice samples using smartphones. Then, we annotated the recorded samples with speaker-wise recording-quality scores and utterance-wise perceived emotion labels. We also benchmark SRC4VC on any-to-any VC, in which we trained a multi-speaker VC model on high-quality speech and used the SRC4VC speakers' voice samples as the source in VC. The results show that the recording quality mismatch between the training and evaluation data significantly degrades the VC performance, which can be improved by applying speech enhancement to the low-quality source speech samples.
\keywords{speech corpus, smartphone-recorded, crowdsourcing, annotation, voice conversion}
\end{abstract}

\vspace{-5pt}
\section{Introduction}\label{sect:intro}
\vspace{-3pt}
Voice conversion (VC) is a technology for transforming the voice characteristics of source speech into those of target speech while keeping the phonetic content of the source speech unchanged~\cite{sisman21}.
VC enables expressions beyond the physical constraints of the human voice and enriches speech communication through social applications, such as speech-to-speech translation~\cite{biadsy19interspeech} and customer service~\cite{okano23vrst}.
As a result of developments in machine learning techniques~\cite{kaneko18cyclegan_vc,kameoka19mcvae,popov22iclr} and well-designed speech corpora~\cite{vctk,zen19}, deep neural network (DNN)-based VC~\cite{desai09nnvc}, which trains a VC model on a speech corpus including multiple speakers' voice samples, has become the mainstream of VC research and improved the quality of converted voices, an essential factor in VC performance evaluation.

The robustness of a trained VC model towards degraded speech input (i.e., degradation robustness~\cite{huang21slt}) is another crucial factor in real-world VC applications.
Namely, model must be capable of high-quality VC even if the source speech, target speech, or both are degraded due to the recording environment and transmission channel. 
The degradation robustness can be improved by using artificially generated noisy speech when training the VC model or by cascading speech enhancement and VC models in the inference~\cite{huang22drvc}.
However, conventional degradation-robust VC work used only artificially degraded speech for VC evaluation.
One primary reason is the lack of publicly available speech corpora recorded by devices owned by end-users with adequate annotations, which hinders validating the degradation robustness of state-of-the-art VC technologies. 


To facilitate research on degradation-robust VC models, we present {\it SRC4VC} (\textbf{S}martphone-\textbf{R}ecorded \textbf{C}orpus for (\textbf{4}) \textbf{V}oice \textbf{C}onversion), a new multi-speaker speech corpus for validating the degradation robustness of a VC model.
The corpus consists of 11 hours of smartphone-recorded speech uttered by 100 crowdsourced Japanese speakers.
Through crowdsourcing, the recorded samples were annotated with speaker-wise recording-quality scores and utterance-wise perceived emotion labels to the recorded samples.
This design makes it possible to evaluate the robustness of a DNN-based VC model towards actual degraded speech input.
We also benchmark SRC4VC on any-to-any VC, in which a multi-speaker VC model is trained on high-quality speech and used the SRC4VC speakers' voice samples as the source speech in VC.
Our contributions are as follows:
    \vspace{-1mm}
    \begin{itemize} \leftskip -1mm \itemsep -0mm
        \item We construct a new, publicly available corpus designed for evaluating the performance of VC from end-users' source speech input. Our corpus is open-sourced for research purposes only from \url{https://y-saito.sakura.ne.jp/sython/Corpus/SRC4VC/index.html}.
        \item We analyze the smartphone-recorded speech samples and discuss the results of speech quality assessments using DNNs and crowdsourcing.
        \item We present the results of an any-to-any VC experiment and show that, for VC using an end-user's voice as the source speech, the recording-quality mismatch can be addressed by simply introducing speech enhancement as preprocessing, rather than training a VC model with data augmentation.        
    \end{itemize}
    \vspace{-2mm}

\vspace{-5pt}
\section{Construction of SRC4VC}
\vspace{-3pt}
\subsection{Core design}\label{subsect:design}
\vspace{-3pt}
We designed SRC4VC so that it includes diverse voice samples from various actual smartphones and advances various VC tasks, such as emotional VC~\cite{zhou22esd} and singing VC~\cite{huang23svcc}.
Specifically, the corpus includes 100 speakers who each recorded 52 voice samples categorized into the following four subsets: 10 read-aloud, 30 expressive, 10 conversational, and two singing samples. 
To examine non-parallel any-to-any VC, we randomly selected sentences for each speaker to record from the subsets except for singing.

{\bf Read-aloud subset:}
We randomly selected 10 sentences per speaker from the Recitation324 subset of ITA~\cite{ita}, which contains 324 phonetically balanced sentences. This subset aims to record neutral, reading-style samples.

{\bf Expressive subset:}
We used JVNV~\cite{xin24jvnv}, which covers six emotions ({\it Angry, Disgust, Fear, Happy, Sadness, Surprise}).
We randomly selected five sentences for each emotion category and asked the speakers to read the presented sentences while expressing the corresponding emotion.

{\bf Conversational subset:}
We used STUDIES~\cite{saito22studies} and CALLS~\cite{saito23calls} consisting of teacher-student and operator-customer dialogues, respectively.
Although these corpora contain human-annotated emotion labels, we did not present the labels to the speakers because the ground-truth emotion for actual conversation generally cannot be defined.
Instead, we allowed the speakers to express their own emotions by interpreting the presented sentences.

{\bf Singing subset:}
We used two Japanese copyright-free songs: ``katatsumuri'' (child-song) and ``Shining star\footnote{\url{https://maou.audio/14_shining_star/} (in Japanese)}'' (J-POP).
We asked the speakers to sing only the first chorus of these two songs because singing entire songs without retakes is difficult for non-professionals.

\vspace{-3pt}
\subsection{Voice recording by crowdworkers}\label{subsect:voice_recording}
\vspace{-3pt}
We crowdsourced smartphone-recorded voice samples using the above-mentioned four subsets.
We collected the samples by macrotask crowdsourcing~\cite{simperl15}, where crowdworkers are employed by clients taking the workers' skills and experiences into consideration.
We describe the recording process in detail below.

{\bf Preparing voice recording platform:}
We created a webpage for voice recording that can be accessed via a smartphone.
The webpage contained instructions for recording for each subset, a recording start/stop button, and text with the pronunciation of the words to be spoken.
The recording function was implemented using Web Audio API\footnote{\url{https://webaudioapi.com/}}.


{\bf Recording speakers' voice samples:}
We recruited speakers through crowdsourcing and required them to understand the purpose of this project (i.e., collecting smartphone-recorded voice samples), have a smartphone, and record their voice in a quiet environment.
We asked the speakers to record their voice using smartphones and our web-based recording platform.
The speakers recorded 52 samples under the four subsets described in Section~\ref{subsect:design} in their own rooms.
They could listen to reference singing voice samples when recording the ``Singing'' subset.
We allowed the speakers to rerecord their voice if they made a mistake in the recording.
After the recording process, the speakers submitted their recordings to a web server owned by the authors.
The sampling rate was 48~kHz.
With our manual validation of recorded data, we asked some of the speakers to rerecord misrecorded samples.


\vspace{-3pt}
\subsection{Annotation}
\vspace{-3pt}
We crowdsourced speaker-wise recording quality scores and utterance-wise perceived emotion labels.

{\bf Speaker-wise recording quality scores:}
We conducted recording quality rating of the collected voice samples and constructed the annotation sets as follows.
Allowing sample overlap among sets, we made 400 sets containing 25 read-aloud utterances for each.
The final speaker-wise recording quality scores were calculated by taking the average of 100 obtained scores for each speaker.

{\bf Utterance-wise perceived emotion labels:}
We conducted emotion classification tests of the recorded ``Expressive'' and ``Conversational'' subsets.
The purpose was to investigate how well the speakers recruited by crowdsourcing were able to express the instructed or situation-oriented emotion in their speech.
For each of the 30 expressive and 10 conversational samples, five different crowdworkers annotated the perceived emotion from seven options (i.e., the six emotions and {\it Neutral}).

\begin{table}[t]
\centering
\caption{Crowdsourcing setup: the numbers of recruited crowdworkers and the amounts of paid rewards (per crowdworker). ``Rec,'' ``Emo,'' ``expr,'' and ``conv'' in this table denote recording, emotion, expressive, and conversational, respectively.}
\vspace{-3mm}
\label{tab:crowdsourcing_setup}
\begin{tabular}{c|cc}
\hline
\hline
Task & \# workers & Rewards \\
\hline
Voice rec. & 100 & \$33.2 \\
Rec quality MOS test & 400 & \$0.44 \\
Emo labeling (expr) & 500 & \$1.31 \\
Emo labeling (conv) & 500 & \$0.44 \\
\hline
\hline
\end{tabular}
\end{table}

\begin{figure}[t]
  \centering
  \includegraphics[width=0.8\linewidth, clip]{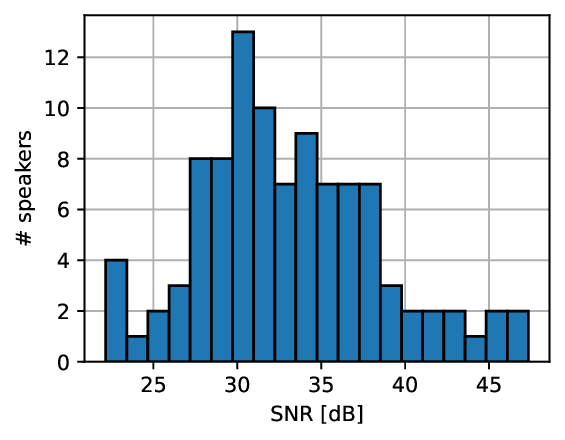}
  \vspace{-8pt}
  \caption{Histogram of SNRs averaged over speakers.}
  \label{fig:snr_hist}
\end{figure}

\begin{figure}[t]
  \centering
  \includegraphics[width=0.8\linewidth, clip]{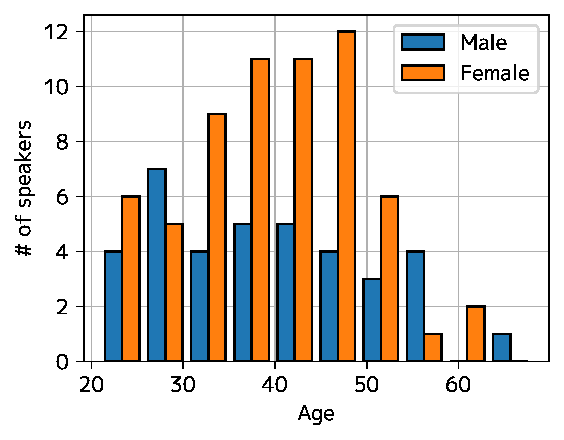}
  \vspace{-8pt}
  \caption{Histogram of speaker ages.}
  \label{fig:age_hist}
\end{figure}

\begin{figure}[t]
  \centering
  \includegraphics[width=0.75\linewidth, clip]{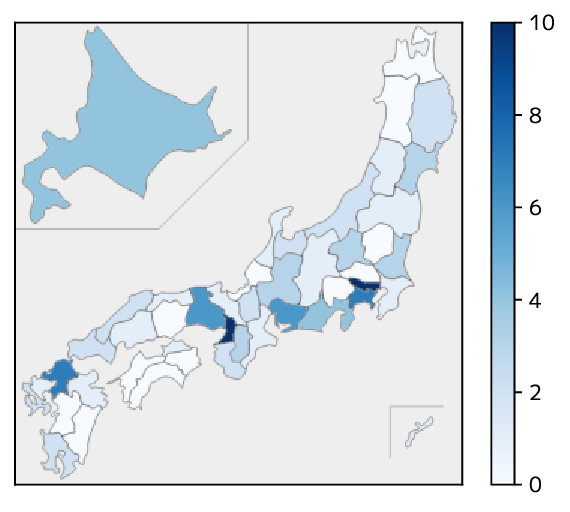}
  \vspace{-8pt}
  \caption{Number of speakers from each prefecture in Japan. The map chart was drawn using the japanmap library.}
  \label{fig:speaker_locale}
\end{figure}

\begin{table*}[t]
\centering
\caption{List of existing open-sourced multi-speaker speech corpora related to SRC4VC. ``Lang.'' and ``Dur.'' denote covered languages and total durations in hours, respectively.}
\vspace{-3mm}
\label{tab:corpora}
\small
\begin{tabular}{cccccc}
\hline
\hline
Corpus & Speaking styles & Lang. & Dur. & \# of speakers & Recording quality \\
\hline
 VCTK~\cite{vctk} & Reading w/ various accents & En & 44 & 109 & Studio-recorded \\
 DDS~\cite{li22dds} & Reading (from VCTK~\cite{vctk} and DAPS~\cite{mysore15daps}) & En & 2000 & 48 & Device-rerecorded \\ \hline 
 JVS~\cite{takamichi20ast} & Reading, Whisper, Falsetto & Ja & 30 & 100 & Studio-recorded \\
 CPJD~\cite{takamichi18cpjd} & Reading w/ dialects & Ja & 7 & 22 & Device-recorded \\
 \textbf{SRC4VC} & Reading, Expressive, Conversational, Singing & Ja & 11 & 100 & Smartphone-recorded \\
\hline
\hline
\end{tabular}
\end{table*}

\vspace{-5pt}
\section{Corpus Analysis}
\vspace{-3pt}
\subsection{Crowdsourcing setting}
\vspace{-3pt}
We used Lancers\footnote{\url{https://www.lancers.jp/}} as the crowdsourcing platform.
The overall period of the voice recording ran 2023.
Table~\ref{tab:crowdsourcing_setup} lists the number of crowdworkers and the amounts of paid rewards per crowdworker.
The recording quality ratings were conducted on the basis of five-scale mean opinion score (MOS) test on a scale of 1 (very bad) to 5 (very good).

\vspace{-3pt}
\subsection{Corpus specification}\label{subsect:spec}
\vspace{-3pt}
Our SRC4VC consists of 1,000 read-aloud, 3,000 expressive, 1,000 conversational, and 200 singing voice samples recorded by 100 speakers, the total durations of which are 1.46, 7.16, 1.66, and 0.87 hours, respectively.

Figure~\ref{fig:snr_hist} shows the histogram of averaged priori signal-to-noise ratio (SNR) per speaker.
We estimated SNRs based on the Ephraim-Malah algorithm~\cite{ephraim84} implemented in Essentia~\cite{bogdanov13}.
The minimum, median, and maximum values of the estimated SNRs with the speaker IDs are 22.2~dB (044), 32.5~dB (017), and 47.3~dB (064), respectively.
These results indicate that our SRC4VC covers wider SNR range (approximately 25~dB) of recorded samples.

\vspace{-3pt}
\subsection{Speaker distribution}\label{subsect:speaker_distribution}
\vspace{-3pt}
We discuss the distributions of SRC4VC speakers regarding gender, age, locale, and used device.
The ratio of male to female speakers was 37 to 63, and the histogram of speaker ages is shown in Figure~\ref{fig:age_hist}.
The youngest and oldest ages with the speaker IDs were 21 (085) and 68 (009), respectively.
Figure~\ref{fig:speaker_locale} maps the speakers' locales.
Our SRC4VC covers 34 of 47 prefectures in Japan as the speaker locales.
Sixty-one percent of the speakers used an iPhone as their recording device, with models ranging from SE to 15 Pro.
These results demonstrate that our SRC4VC corpus consists of speech recorded by diverse speakers on various smartphones.

\vspace{-3pt}
\subsection{Comparison with existing corpora}
\vspace{-3pt}
Table~\ref{tab:corpora} lists existing open-sourced multi-speaker speech corpora related to SRC4VC.
Although SRC4VC is smaller than existing de-facto standard corpora for developing high-quality multi-speaker VC technologies, such as VCTK~\cite{vctk} and JVS~\cite{takamichi20ast}, it covers various speech degradations that rarely occur in controlled and supervised studio recording.
DDS~\cite{li22dds} includes aligned parallel high-quality speech recordings taken from VCTK~\cite{vctk} and DAPS~\cite{mysore15daps} and their rerecorded versions with various conditions by combining diverse acoustic environments and devices.
In contrast, SRC4VC contains actual end-users' voice samples recorded in their own acoustic environments using smartphones. 
The design of SRC4VC was inspired by that of CPJD~\cite{takamichi18cpjd}, in which the developers collected parallel speech data of Japanese dialects by crowdsourcing.
As discussed in Section~\ref{subsect:speaker_distribution}, our SRC4VC covers speakers from a wider area than those of CPJD (21 dialects).

\begin{table}[t]
\centering
\caption{SRCC between human-annotated recording quality and each NISQA score. ``Nat.'' denotes Naturalness.}
\vspace{-3mm}
\label{tab:srcc_mos}
\footnotesize
\begin{tabular}{cccc|c}
\hline
Noisiness & Coloration & Discontinuity & Loudness & Nat. \\
\hline
 0.15 & \textbf{0.67} & \textbf{0.62} & 0.36 & 0.54 \\
\hline
\end{tabular}
\end{table}

\begin{table}[t]
\centering
\caption{Percentages of agreed emotional utterances in ``Expressive'' (expr) and ``Conversational'' (conv) subsets.}
\vspace{-3mm}
\label{tab:emo}
\footnotesize
\begin{tabular}{c|ccccccc}
\hline
Subset & Ang & Dis & Fea & Hap & Sad & Sur & Neu \\
\hline
expr & 14.6 & 17.8 & 14.4 & 16.7 & 15.7 & 17.3 & 0.35 \\
conv & 0.45 & 0.29 & 0.08 & 0.59 & 0.56 & 0.28 & 1.08 \\
\hline
\end{tabular}
\end{table}

\vspace{-3pt}
\subsection{Annotation results}\label{subsect:annotation_results}
\vspace{-3pt}
We analyze the results of the annotations for recording quality and emotion labels.

{\bf Recording quality scores:}
We computed the Spearman's rank correlation coefficient (SRCC) between the human-annotated recording quality MOS values and NISQA scores~\cite{mittag21nisqa}, which quantify the quality of input speech in terms of {\it Noisiness, Coloration, Discontinuity, and Loudness}~\cite{waltermann12,cote07}, in addition to {\it Naturalness}.
Table~\ref{tab:srcc_mos} lists the results.
The ``Coloration'' and ``Discontinuity'' scores moderately correlated with the human-annotated MOS values.
These results indicate that dominant speech degredation factors in SRC4VC are frequency response and non-linear distortion.

{\bf Emotion labels:}
We investigated the inter-annotator agreement of the perceived emotion labels.
Specifically, we calculated the percentages of ``agreed emotional utterances.''
Namely, if more than two annotators assigned the same perceived emotion label to one utterance, we regarded it as the agreed emotional utterance.
Table~\ref{tab:emo} lists the results.
The distribution of agreed emotion labels for the ``Expressive'' subset was close to uniform except for ``Neutral.''
In contrast, the percentage of agreed emotion labels for the ``Conversational'' subset was generally low, with the highest percentage of 1\% for the ``Neutral'' emotion.
These results suggest that the SRC4VC speakers could express and utter the indicated emotion to a certain extent, though they were not necessarily professional actors.

\vspace{-5pt}
\section{Any-to-Any VC Experiment}
\vspace{-3pt}

\subsection{Experimental conditions}
\vspace{-3pt}
Our experiment aims to investigate whether the recording-quality mismatch between the training data and evaluation data actually worsens VC performance and what kind of approaches can mitigate the mismatch.

{\bf Datasets:}
We used JVS~\cite{takamichi20ast} for building an any-to-any VC model.
The training, validation, and test sets included 86 (jvs001--jvs086), 4 (jvs087--jvs090), and 10 (jvs091--jvs100) speakers, respectively.
We used the parallel100 subset of JVS including 100 parallel speech utterances for each speaker as the training and validation sets.
The test set, which was used as reference speech samples of unseen target speakers in any-to-any VC, was the nonpara30 subset of JVS including 30 non-parallel speech utterances for each speaker.
Therefore, the trained VC model was evaluated on the task of converting 100 SRC4VC speakers' samples into those of the 4 JVS test speakers.
Although our SRC4VC corpus contains the ``Singing'' subset, we excluded it from the evaluation data.
The reason was that the VC task, i.e., zero-shot singing VC using a model trained on reading-style speech only, was too difficult for the trained model.
We downsampled all speech data to 16~kHz.

{\bf Backbone VC model:}
We adopted S2VC~\cite{lin21s2vc} as the backbone VC model for our experiment.
The S2VC model incorporates self-supervised learning features~\cite{huang22s3prl_vc} as the intermediate representations used for VC.
Referring to Huang et al.'s degradation-robust VC study~\cite{huang22drvc}, we trained the S2VC model to generate target speaker's mel-spectrograms from the source and target speakers' contrastive predictive coding features~\cite{oord18cpc}.
We used the implementation of S2VC publicly available in GitHub\footnote{\url{https://github.com/cyhuang-tw/robust-vc/tree/main}}.
We trained S2VC with 100K iterations and used the model that achieved the lowest validation loss for our evaluation.
The initial learning rate was set to 5.0$\times$10$^{-5}$ and decayed with warm-up cosine annealing~\cite{loshchilov17}.
The optimizer was AdamW~\cite{loshchilov19}.
We also trained HiFi-GAN~\cite{kong20} to synthesize a speech waveform from a mel-spectrogram using the same training data for S2VC.
The optimization algorithm was Adam~\cite{kingma14adam} with $\eta$ of 0.0002, $\beta_1$ of 0.8, and $\beta_2$ of 0.99.
We trained HiFi-GAN with 600K iterations.
The computational resources were two NVIDIA GeForce 1080 Ti GPUs.

{\bf Compared methods:}
We compared data augmentation and preprocessing techniques for improving the robustness of VC~\cite{huang22drvc}.
Specifically, we examined three data augmentation methods: additive noise, reverberation, and band rejection, using the same configurations as those in Huang et al.'s study~\cite{huang22drvc}.
We also used two speech enhancement models: Demucs~\cite{defossez20}\footnote{\url{https://github.com/facebookresearch/demucs}} and Miipher~\cite{koizumi23miipher})\footnote{\url{https://github.com/Wataru-Nakata/miipher}} with their official and unofficial implementation available on GitHub, respectively.
For simplicity, we did not investigate combinations of these techniques, i.e., applying two or more data augmentations combined with speech enhancement.

{\bf Speaker groups:}
To investigate the relation between recording quality and VC performance, we first divided the 100 SRC4VC speakers into four groups (Q1--Q4) using the quantile values of their recording-quality MOS.
Then, we randomly sampled 250 samples from each group.
Finally, we paired the samples with randomly selected JVS test speakers' voice and used them as the source and target in VC.

\begin{table}[t]
\centering
\caption{Objective evaluation results (NISQA Naturalness). ``DA'' and ``SE'' denote data augmentation and speech enhancement, respectively.}
\vspace{-3mm}
\label{tab:obj}
\footnotesize
\begin{tabular}{l|ccccc|c}
\hline
        & \multicolumn{5}{c|}{SRC4VC} & \\ 
Methods & Q1 & Q2 & Q3 & Q4 & Avg. & JVS \\
\hline
Baseline (B) & 2.36 & 2.50 & 2.38 & 2.41 & 2.41 & 2.83 \\
B + DA (noise) & 2.39 & 2.53 & 2.41 & 2.47 & 2.45 & 2.82 \\
B + DA (reverb) & 2.33 & 2.46 & 2.48 & 2.50 & 2.44 & 2.90 \\
B + DA (band) & 2.34 & 2.51 & 2.43 & 2.51 & 2.45 & 2.82 \\
\hline
B + SE (Demucs) & 2.39 & 2.50 & 2.42 & 2.43 & 2.43 & N/A \\
B + SE (Miipher) & 2.49 & 2.60 & 2.51 & 2.50 & \textbf{2.52} & N/A \\
\hline
\end{tabular}
\end{table}

\vspace{-3pt}
\subsection{Objective evaluation}
\vspace{-3pt}
As the objective evaluation criteria, we used the NISQA {\it Naturalness} score~\cite{mittag21nisqa} of the converted speech.
We did not evaluate the JVS-to-JVS VC cases when we used speech enhancement as the preprocessing.

Table~\ref{tab:obj} shows the evaluation results.
The predicted \textit{Naturalness} scores of the two VC cases using JVS and SRC4VC as the source speakers were significantly different, indicating that the recording-quality mismatch between the training and evaluation actually worsened the naturalness of converted speech.
The introduced data augmentation strategies could mitigate the deterioration in naturalness, but the effectiveness was not so meaningful.
We observed that the relatively high-quality recorded speakers (i.e., Q3 and Q4) benefited from the data augmentation, while Q1 and Q2 suffered from the naturalness degradation when inappropriate data augmentations for them (i.e., reverb or band) were adopted.
The optimal strategy was using Miipher for speech enhancement.
These results suggest that, for VC using an end-user's voice as the source speech, the recording-quality mismatch can be addressed by simply introducing speech enhancement as preprocessing, rather than training a VC model with data augmentation.

\vspace{-3pt}
\subsection{Subjective evaluation}
\vspace{-3pt}
We conducted two five-scaled MOS tests regarding the naturalness and speaker similarity of converted speech.
In the naturalness test, we presented 30 converted speech samples to listeners in random order.
The listeners rated the naturalness of each speech sample with an integer between 1 (very bad) to 5 (very good).
In the similarity test, listeners first listened to the target speaker's natural speech for reference.
Then, they scored how similar the speaker of the presented voice was to the reference.
A total of 30 synthetic speech samples were presented.
Two hundred listeners participated in each test using our crowdsourcing evaluation system, for a total of 400 listeners.
Unlike the objective evaluation, we did not evaluate JVS-to-JVS converted samples or distinguish the SRC4VC source speakers by the recording-quality scores.

\begin{table}[t]
\centering
\caption{Results of naturalness and similarity MOS tests.}
\vspace{-3mm}
\label{tab:sbj}
\begin{tabular}{l|cc}
\hline
Methods & Naturalness & Similarity \\ 
\hline
Baseline (B) & 2.54 & 2.17 \\
B + DA (noise) & 2.59 & 2.18 \\
B + DA (reverb) & 2.66 & \textbf{2.22} \\
B + DA (band) & 2.62 & 2.17 \\
\hline
B + SE (Demucs) & 2.53 & 2.17 \\
B + SE (Miipher) & \textbf{2.74} & 2.21 \\
\hline
\end{tabular}
\end{table}

Table~\ref{tab:sbj} shows the results of the subjective evaluation.
The ``Baseline + SE (Miipher)'' achieved the highest naturalness, whose score was statistically significant compared to that of ``Baseline'' ($p < 0.05$).
This further demonstrates the effectiveness of speech enhancement preprocessing.
From this result, we decided to publish the restored version of SRC4VC (i.e., {\it SRC4VC-R}) as well.
Focusing on the evaluation results of speaker similarity, there was no statistical significance among the obtained scores, although the score of ``Baseline + DA (reverb)'' was slightly higher than the others.

\vspace{-5pt}
\section{Conclusion}
\vspace{-3pt}
We presented SRC4VC, a new multi-speaker speech corpus for validating the degradation robustness of a VC model.
The results of any-to-any VC experiments demonstrated that the recording-quality mismatch between the training and evaluation data significantly worsened the VC performance, and applying speech enhancement to the low-quality source speech samples improved the performance.
In the future, we intend to develop a method to adapt a VC model trained with clean speech to an end-user's degraded speech input.
We also continue to enlarge SRC4VC so that it includes speech degradations caused by transmission channels (e.g., live VoIP calls covered by the test subsets of NISQA corpus~\cite{mittag21nisqa}).

\textbf{Acknowledgements:}
This research was conducted as joint research between LY Corporation and Saruwatari-Takamichi Laboratory of The University of Tokyo, Japan. This work was supported by Research Grant S of the Tateishi Science and Technology Foundation.

   \ninept
   \bibliographystyle{IEEEtran}
   \bibliography{template}

\end{document}